\documentclass[superscriptaddress,secnumarabic,amssymb,amsmath,nobibnotes,aps,prd,showkeys,showpacs,nofootinbib]{revtex4}%
\usepackage{graphicx}
\usepackage{epsf}
\usepackage{bm}
\usepackage{amsmath}
\usepackage{amsfonts}
\usepackage{amssymb}%
\providecommand{\U}[1]{\protect\rule{.1in}{.1in}}
\newcommand{\be}{\begin{equation}}
\newcommand{\ee}{\end{equation}}

\newcommand{\mincir}{\raise
-3.truept\hbox{\rlap{\hbox{$\sim$}}\raise4.truept\hbox{$<$}\ }}
\newcommand{\magcir}{\raise
-3.truept\hbox{\rlap{\hbox{$\sim$}}\raise4.truept\hbox{$>$}\ }}

\begin{document}

\title{Noether symmetries and analytical solutions in $f(T)$-cosmology: A complete study}

\author{S. Basilakos}
\affiliation{Academy of Athens, Research Center for Astronomy and Applied Mathematics,
Soranou Efesiou 4, 11527, Athens, Greece}

\author{S. Capozziello}
\affiliation{Dipartimento di Fisica, Universita' di Napoli "Federico II}
\affiliation{and INFN Sez. di Napoli, Compl. Univ. di Monte S. Angelo, Ed. G., Via Cinthia,
9, I-80126, Napoli, Italy.}

\author{M. De Laurentis}
\affiliation{Dipartimento di Fisica, Universita' di Napoli "Federico II}
\affiliation{and INFN Sez. di Napoli, Compl. Univ. di Monte S. Angelo, Ed. G., Via Cinthia,
9, I-80126, Napoli, Italy.}

\author{A. Paliathanasis}
\affiliation{Faculty of Physics, Department of Astrophysics - Astronomy - Mechanics
University of Athens, Panepistemiopolis, Athens 157 83, Greece}

\author{M. Tsamparlis}
\affiliation{Faculty of Physics, Department of Astrophysics - Astronomy - Mechanics
University of Athens, Panepistemiopolis, Athens 157 83, Greece}

\begin{abstract}
We investigate the main features of the 
flat Friedmann-Lema\^\i tre-Robertson-Walker cosmological models 
in the $f(T)$ modified gravity regime. In particular,
a general approach to find out 
exact cosmological solutions in $f(T)$
gravity is discussed. Instead of taking into account phenomenological
models, we consider as a selection criterion, the existence of Noether
symmetries in the cosmological $f(T)$ point-like Lagrangian.
We find that only the $f(T)=f_{0}T^{n}$ model admits 
extra Noether symmetries.
The existence of extra
Noether integrals can be used in order 
to simplify the system of differential equations
(equations of motion) as well as to determine the integrability of the 
$f(T)=f_{0}T^{n}$ cosmological model. 
Within this context,
we can solve the problem analytically and 
thus we provide the
evolution of the main cosmological functions such as the scale factor of the
universe, the Hubble expansion rate, the deceleration parameter and 
the linear matter perturbations.
We show that the $f(T)=f_{0}T^{n}$ cosmological model suffers from two 
basic problems. The first problem is related to the fact that 
the deceleration parameter is constant which means that it 
never changes sign, and therefore the
universe always accelerates or always decelerates
depending on the value of $n$. 
Secondly, we find that 
the clustering  growth rate  remains always equal to unity 
implying that the recent growth data disfavor the 
$f(T)=f_{0}T^{n}$ gravity. 
Finally, we prove that the 
$f(T)=f_{0}T^{n}$ gravity can be cosmologically equivalent with the 
$f(R)=R^{n}$ gravity model and the time 
varying vacuum model $\Lambda(H)=3\gamma H^{2}$ (for $n^{-1}=1-\gamma$)  
because the above cosmological scenarios share 
exactly the same Hubble expansion, 
despite the fact that the three models 
have a different geometrical origin.
Finally, some important differences with 
power-law $f(R)$-gravity are pointed out.
\end{abstract}
\pacs{98.80.-k, 95.35.+d, 95.36.+x}
\keywords{Cosmology; dark energy; alternative gravity theories; torsion; exact solutions..}\maketitle

\section{Introduction}
Non-standard gravity models provide an alternative possibility 
towards understanding the accelerated expansion of the Universe
(see \cite{Teg04} and references therein). 
The physical mechanism which is responsible 
for the present accelerating stage of the universe
can be driven by a modification of the 
Einstein-Hilbert action, while the matter content of the universe remains
the same (relativistic and cold dark matter). 
In the literature there are plenty of modified gravity 
models proposed by different authors, such as the braneworld 
Dvali, Gabadadze and Porrati \cite{Dvali2000} model, 
$f(R)$ gravity \cite{Sot10}, 
scalar-tensor theories \cite{scal}, Gauss-Bonnet gravity \cite{gauss}, 
Ho\v{r}ava-Lifshitz gravity \cite{hor}, nonlinear massive gravity 
\cite{Hinterbichler:2011tt} etc. 

Another gravitational scenario which has recently gained a lot of attention
is the so called $f(T)$ gravity. 
The intrinsic properties of this scenario are based 
on the rather old formulation of the teleparallel equivalent of
General Relativity (TEGR) \cite{ein28,Hayashi79,Maluf:1994ji}. 
Specifically, instead of using the torsion-less Levi-Civita connection
of the classical General Relativity (GR) one utilizes 
the curvature-less Weitzenb{\"{o}}ck connection in which the 
corresponding dynamical fields are the four linearly
independent {\it vierbeins.}
Therefore, all the information concerning the gravitational field
are included in the torsion tensor. Within this framework,  
considering invariance under general
coordinate transformations, global Lorentz-parity transformations, and
requiring up to second order terms of the torsion tensor, 
one can write down the corresponding Lagrangian
density $T$ \cite{Hayashi79} by using some 
suitable contractions. A natural generalization of  TEGR gravity 
is  $f(T)$ gravity which is based on the fact that
we allow the
Lagrangian to be a function of $T$ \cite{Ferraro,Ferraro:2006jd,Linder:2010py},
inspired, of course, by the well-known extension of $f(R)$ 
Einstein-Hilbert action. However,  
$f(T)$ gravity does not coincide with $f(R)$ extension, but it rather
consists a different class of modified gravity.
It is interesting to mention 
that the torsion tensor includes only products of
first derivatives of the vierbeins, giving rise to 
second-order field differential equations
in contrast with the $f(R)$ gravity that provides fourth-order 
equations which potentially may lead to some problems, for example in the well-position and well-formulation of Cauchy problem \cite{vignolo}.  

Despite the fact that TEGR coincides completely with GR,
both at the background and perturbation levels, 
it has been shown that $f(T)$ gravity provides different 
structural properties with respect to GR as well as different
black-hole solutions and cosmological features \cite{Ferraro,Ferraro:2006jd,Linder:2010py,Myrzakulov:2010vz,
Wu:2010mn,Bengochea001,Iorio:2012cm,Wang:2011xf}. 
An important question here is what classes of $f(T)$ 
extensions are allowed. From the 
phenomenological viewpoint, the aforementioned 
cosmological and spherical analysis lead to a variety of such 
expressions. Using cosmological \cite{Wu:2010mn,Bengochea001,Zhang:2012jsa} and Solar System \cite{Iorio:2012cm}
observations, one can show that the deviations from TEGR must be small.

In this work, we use a model-independent selection rule based
on first integrals, due to Noether symmetries of the equations of
motion, in order to identify the viability of $f(T)$ gravity in the context of
flat  Friedmann-Lema\^\i tre-Robertson-Walker  (FLRW) cosmologies. 
Actually, the idea to use Noether symmetries
in cosmology is not new and indeed there is a lot of work 
in the literature (see
\cite{Cap96,Cap97,RubanoSFQ,Sanyal05,Szy06,Cap07,Capa07,Bona07,Cap08,Cap09,Vakili08,Yi09,WeiHao}) along this line. In this context, recently we have shown  
(see Basilakos et al. \cite{Basilakos11}; Paliathanasis 
et al. \cite{Tsafr}) that
the existence of Noether symmetries can be used as a selection
criterion in order to distinguish the scalar dark energy models \cite{Basilakos11} as well as the $f(R)$ gravity models \cite{Tsafr}.
Inspired by the above works in the current article, 
we would like to estimate the 
Noether symmetries of the $f(T)$ gravity.
The aim here is $(a)$ to identify the $f(T)$ functional forms which 
accommodate extra Noether symmetries, and
$(b)$ for these models, to solve the system of the resulting field equations and
derive analytically the main cosmological
functions (the scale factor, the Hubble expansion rate, deceleration
parameter and growth factor) and 
finally to compare with other cosmological patterns which 
are outside and inside GR. 

The structure of the article is as follows: 
In Sec. II, we discuss the issue of 
torsion in GR and its connection with 
unholonomic frames. This discussion is useful  in order to 
clarify some misunderstandings on the role of torsion 
that are present in literature. In particular, we shall 
discuss its dependence on the frame where observations are made.

In Sec. III, we give 
the basic FLRW
cosmological equations in the framework of $f(T)$ gravity.
The main properties and theorems of the Noether Symmetry Approach
are summarized
in Sec. IV.  Noether symmetries for  
$f(T)$ cosmology are discussed in Sec. V. 
In Sec VI we provide analytical solutions for 
$f(T)$ models that admit non trivial Noether symmetries. A comparison with analogous $f(R)$ cosmology is pursued putting in evidence similarities and differences. We draw  conclusions in Sec.VI.

\section{The role of  torsion in General Relativity}
Before starting our considerations on $f(T)$ gravity and its cosmological realization, it is useful to discuss in detail the role of torsion in GR considering, in particular, how it behaves with respect to holonomic and unholonomic frames.

Let us start with some definitions.
In an $n-$dimensional manifold ${\cal M}$ consider a coordinate neighborhood ${\cal U}$
with a coordinate system $\{x^{\mu }\}.$ At each point $P \in {\cal U}$, we have
the resulting holonomic frame $\{\partial _{\mu }\}.$ We define in ${\cal U}$ a
new frame $\{e_{a}(x^{\mu })\}$ which is related to the holonomic frame $%
\{\partial _{a}\}$ as follows:%
\begin{equation}
e_{a}(x^{\mu })=h_{a}^{\mu }\partial _{\mu }~\ \ a,\mu =1,2,...,n
\label{H.1}
\end{equation}%
where the quantities $h_{a}^{\mu }(x)$ are in general functions of the
coordinates (i.e. depend on the point $P$). Notice that Latin indexes count
vectors, while Greek indexes are tensor indices. We assume that $\det
h_{a}^{\mu }\neq 0$ which guaranties that the vectors $\{e_{a}(x^{\mu })\}$
form a set of linearly independent vectors. We define the ''inverse''
quantities $h_{a }^{\mu}$ by means of the following ''orthogonality''
relations:
\begin{equation}
h_{a}^{\mu }h_{\nu }^{a}=\delta _{\nu }^{\mu }\;\;\;,h_{b}^{\mu }h_{\mu
}^{c}=\delta _{b}^{c}.  \label{H.1a}
\end{equation}
The commutators of the vectors $\{e_{a}\}$ are not in general all zero. If
they are zero, then there exists a new coordinate system in ${\cal U}$,  $\{y^{b}\}$
 so that ${\displaystyle e_{b}=\frac{\partial }{\partial y^{b}}}$, i.e. the new frame is
holonomic. If there are commutators $[e_{a},e_{b}]\neq 0$ then the new frame
$\{e_{b}\}$ is called unholonomic and at least a number of  vectors $e_{b}$ cannot be written
in the form $e_{b}=\partial _{b}.$ The quantities which characterize an
unholonomic frame are the objects of unholonomicity or Ricci rotation
coefficients $\Omega _{\text{ }bc}^{a}$ defined by the relation%
\begin{equation}
\lbrack e_{a},e_{b}]=\Omega _{\text{ }ab}^{c}e_{c}.  \label{H.3}
\end{equation}
Let us  compute:%
\begin{equation*}
\lbrack e_{a},e_{b}]=[h_{a}^{\mu }\partial _{\mu },h_{b}^{\nu }\partial
_{\nu }]=\left[ h_{a}^{\mu }h_{b,\mu }^{\nu }h_{\nu }^{c}-h_{b}^{\nu
}h_{a,\nu }^{\mu }h_{\mu }^{c}\right] e_{c}
\end{equation*}%
from which follows that the Ricci rotation coefficients of the frame $%
\{e_{a}\}$ are:
\begin{equation}
\Omega _{\text{ }bc}^{a}=2h_{[b}^{\mu }h_{c],\mu }^{\nu }h_{\nu }^{a}.
\label{H.2}
\end{equation}%
The condition for $\{e_{a}\}$ to be a holonomic basis is $\Omega _{\text{ }%
bc}^{a}=0$ at all points $P\in  {\cal U}.$ This is a set of linear partial
differential equations whose solution defines all holonomic frames and all
coordinate systems in ${\cal U}.$ One obvious solution is $h_{b}^{c}=\delta _{b}^{c}
$. The set of all coordinate systems in ${\cal U}$, equipped with the operation of
composition of transformations, has the structure of an infinite dimensional
Lie group which is called the {\it Manifold Mapping Group}
\cite{StephaniB}. 

Let us  consider now the special unholonomic frames which satisfy the Jacobi
identity:%
\begin{equation}
\lbrack \lbrack
e_{a},e_{b}],e_{c}]+[[e_{b},e_{c}],e_{a}]+[[e_{c},e_{a}],e_{b}]=0.
\label{H.4}
\end{equation}%
These frames are the generators of a Lie algebra, therefore they have an
extra role to play. Replacing the commutator in terms of the unholonomicity
objects, we find the following identity:%
\begin{equation}
\Omega _{\text{ }ab,c}^{d}+\Omega _{\text{ }ba,a}^{d}+\Omega _{\text{ }%
ca,b}^{d}-\Omega _{\text{ }ab}^{l}\Omega _{cl}^{d}-\Omega _{\text{ }%
bc}^{l}\Omega _{al}^{d}-\Omega _{\text{ }ca}^{l}\Omega _{bl}^{d}=0.
\end{equation}
Using the definition of the covariant derivative we write:%
\begin{equation}
\nabla _{e_{i}}e_{j}=\Gamma _{ij}^{k}e_{k}  \label{H.20}
\end{equation}%
where $\Gamma _{ij}^{k}$ are the connection coefficients in the frame $%
\{e_{i}\}.$ If we compute the $\Gamma _{ij}^{k}$ assuming 
\begin{equation*}
\lbrack e_{i},e_{j}]=C_{.ij}^{k}e_{k}
\end{equation*}%
it follows that%
\begin{equation*}
C_{.ij}^{k}=\Omega _{.jk}^{k}.
\end{equation*}

Let us consider now  three vector fields $X,Y,Z$ and the covariant 
derivative of the
metric vector $X.$ Then we have:

\begin{equation}
\nabla _{X}g(Y,Z)=X(g(Y,Z))-g(\nabla _{X}Y,Z)-g(Y,\nabla _{Y}Z)  \label{H.21}
\end{equation}%
and by interchanging the role of $X,Y,Z:$%
\begin{equation}
\nabla _{Y}g(Z,X)=Y(g(Z,X))-g(\nabla _{Y}Z,X)-g(Z,\nabla _{Z}X)  \label{H.22}
\end{equation}

\begin{equation}
\nabla _{Z}g(X,Y)=Z(g(X,Y))-g(\nabla _{Z}X,Y)-g(X,\nabla _{X}Y).
\label{H.23}
\end{equation}

Adding Eqs. (\ref{H.21}),  (\ref{H.22}) and subtracting (\ref{H.23}),  one obtains:

\begin{eqnarray*}
\nabla _{X}g(Y,Z)+\nabla _{Y}g(Z,X)-\nabla _{Z}g(X,Y)
&=&X(g(Y,Z))+Y(g(Z,X))-Z(g(X,Y))+ \\
&&-\left[ g(\nabla _{X}Y,Z)+g(\nabla _{Y}Z,X)-g(\nabla _{Z}X,Y)\right] + \\
&&-\left[ g(Y,\nabla _{X}Z)+g(Z,\nabla _{Y}X)-g(X,\nabla _{Z}Y)\right]
\end{eqnarray*}%
then%
\begin{eqnarray*}
\nabla _{X}g(Y,Z)+\nabla _{Y}g(Z,X)-\nabla _{Z}g(X,Y)
&=&X(g(Y,Z))+Y(g(Z,X))-Z(g(X,Y))+ \\
&&-\left[ g(\nabla _{X}Y,Z)+g(Z,\nabla _{Y}X)\right] + \\
&&-\left[ g(\nabla _{Y}Z,X)-g(X,\nabla _{Z}Y)\right] + \\
&&-\left[ g(Y,\nabla _{X}Z)-g(\nabla _{Z}X,Y)\right] 
\end{eqnarray*}%
that is%
\begin{eqnarray*}
\nabla _{X}g(Y,Z)+\nabla _{Y}g(Z,X)-\nabla _{Z}g(X,Y) 
&=&X(g(Y,Z))+Y(g(Z,X))-Z(g(X,Y))+ \\
&&-\left[ g(Z,\nabla _{X}Y+\nabla _{Y}X)+g(X,\nabla _{Y}Z-\nabla
_{Z}Y)+g(Y,\nabla _{X}Z-\nabla _{Z}X)\right] \,,
\end{eqnarray*}%
where
\begin{equation*}
g(Z,\nabla _{X}Y+\nabla _{Y}X)=2g\left( Z,\nabla _{X}Y\right) +g\left(
Z,\nabla _{Y}X-\nabla _{X}Y\right) .
\end{equation*}%
Replacing in the last relation and solving for $2g\left( Z,\nabla
_{X}Y\right) $,  we find%
\begin{eqnarray*}
2g\left( Z,\nabla _{X}Y\right)  &=&\left[ X(g(Y,Z))+Y(g(Z,X))-Z(g(X,Y))%
\right] + \\
&&-\left[ \nabla _{X}g(Y,Z)+\nabla _{Y}g(Z,X)-\nabla _{Z}g(X,Y)\right] + \\
&&-\left[ g\left( Z,\nabla _{Y}X-\nabla _{X}Y\right) +g(X,\nabla
_{Y}Z-\nabla _{Z}Y)+g(Y,\nabla _{X}Z-\nabla _{Z}X)\right]
\end{eqnarray*}%
or%
\begin{eqnarray*}
2g\left( Z,\nabla _{X}Y\right)  &=&\left[ X(g(Y,Z))+Y(g(Z,X))-Z(g(X,Y))%
\right] + \\
&&-\left[ \nabla _{X}g(Y,Z)+\nabla _{Y}g(Z,X)-\nabla _{Z}g(X,Y)\right] + \\
&&-\left[ g\left( Z,\nabla _{Y}X-\nabla _{X}Y-\left[ Y,X\right] \right)
+g(X,\nabla _{Y}Z-\nabla _{Z}Y-\left[ Y,Z\right] )+g(Y,\nabla _{X}Z-\nabla
_{Z}X-\left[ X,Z\right] )\right] + \\
&&-\left[ g\left( Z,\left[ Y,X\right] \right) +g\left( X,\left[ Y,Z\right]
\right) +g\left( Y,\left[ X,Z\right] \right) \right] .
\end{eqnarray*}%
At this point, we can define the quantities%
\begin{eqnarray*}
T_{\nabla }(X,Y) = \nabla _{X}Y-\nabla _{Y}X-[X,Y]\,,\qquad
A_{\nabla }(X,Y,Z) = \nabla _{X}g(Y,Z)\,.
\end{eqnarray*}%
The tensors $T_{\nabla }$ and $A_{\nabla }$ are called the {\it torsion}
($T_{\nabla }\equiv T$) and the {\it metricity} of 
the connection $\nabla $ respectively. Last
relation in terms of the fields $T_{\nabla }$ and $A_{\nabla }$ is written
as follows$:$%
\begin{eqnarray}
2g\left( Z,\nabla _{X}Y\right)  &=&\left[ X(g(Y,Z))+Y(g(Z,X))-Z(g(X,Y))%
\right] +  \notag \\
&&-\left[ A_{\nabla }\left( X,Y,Z\right) +A_{\nabla }\left( Y,Z,X\right)
-A_{\nabla }\left( Z,X,Y\right) \right] +  \notag \\
&&-\left[ g\left( Z,T_{\nabla }\left( Y,X\right) \right) +g\left(
X,T_{\nabla }\left( Y,Z\right) \right) +g\left( Y,T_{\nabla }\left(
X,Z\right) \right) \right]   \notag \\
&&-\left[ g\left( Z,\left[ Y,X\right] \right) +g\left( X,\left[ Y,Z\right]
\right) +g\left( Y,\left[ X,Z\right] \right) \right] .  \label{H.24}
\end{eqnarray}
Let $X=e_{l}~,~Y=e_{j}$ and $Z=e_{k}.$ Contracting with $\frac{1}{2}g^{il}$,
we have%
\begin{equation*}
2g\left( Z,\nabla _{X}Y\right) \rightarrow \Gamma _{jk}^{i}
\end{equation*}%
\begin{equation*}
\left[ X(g(Y,Z))+Y(g(Z,X))-Z(g(X,Y))\right] \rightarrow \left\{
_{jk}^{i}\right\}
\end{equation*}%
\begin{equation*}
g\left( X,T_{\nabla }\left( Y,Z\right) \right) \rightarrow Q_{.kj}^{i}
\end{equation*}%
\begin{equation*}
g\left( Z,T_{\nabla }\left( Y,X\right) \right) +g\left( Y,T_{\nabla }\left(
X,Z\right) \right) \rightarrow g^{il}(g_{tj}Q_{kl}^{t}+g_{tk}Q_{jl}^{t})=-%
\bar{S}_{.kj}^{i}
\end{equation*}%
\begin{equation*}
g\left( X,\left[ Y,Z\right] \right) \rightarrow \frac{1}{2}C_{.jk}^{i}
\end{equation*}%
\begin{equation*}
g\left( Z,\left[ Y,X\right] \right) +g\left( Y,\left[ X,Z\right] \right) =%
\frac{1}{2}g^{il}(g_{tj}C_{lk}^{t}+g_{tk}C_{jl}^{t})=-S_{.kj}^{i}
\end{equation*}%
and%
\begin{equation*}
A_{\nabla }\left( X,Y,Z\right) +A_{\nabla }\left( Y,Z,X\right) -A_{\nabla
}\left( Z,X,Y\right) \rightarrow \frac{1}{2}g^{il}\Delta _{jkl}
\end{equation*}%
Replacing in Eq.(\ref{H.24}),  we find the connection coefficients in the frame $%
\{e_{i}\}$, that is 
\begin{equation}
\Gamma _{jk}^{i}=\left\{ _{jk}^{i}\right\} +\bar{S}_{.kj}^{i}+S_{.kj}^{i}-%
\frac{1}{2}g^{il}\Delta _{jkl}+Q_{jk}^{i}-\frac{1}{2}C_{.jk}^{i}  \label{H25}
\end{equation}%
where $\left\{ _{jk}^{i}\right\} $ are  the standard Levi-Civita connection
coefficients (i.e. the Christofell symbols). 
This is the most general expression
for the connection coefficients in terms of the fields $\left\{
_{jk}^{i}\right\} ,$ $T_{\nabla },$ $A_{\nabla }$ and $C_{jk}^{i}$.
Concerning the symmetric and antisymmetric part, we have:%
\begin{eqnarray}
\Gamma _{.(jk)}^{i} &=&\left\{ _{jk}^{i}\right\} +\bar{S}%
_{.jk}^{i}+S_{.jk}^{i}-\frac{1}{2}g^{il}\Delta _{jkl}  \label{H.26} \\
\Gamma _{.[jk]}^{i} &=&Q_{.jk}^{i}-\frac{1}{2}C_{.jk}^{i}\,, \label{H.27}
\end{eqnarray}

and then we can draw the following conclusions:
\begin{enumerate}

\item The connection coefficients in a frame $\{e_{i}\}$ are determined from
the metric, the torsion, the metricity and the unholonomicity objects
(equivalently the commutators) of the frame.

\item  The symmetric part $\Gamma _{.(jk)}^{i}$ of $\Gamma _{jk}^{i}$ depends on
all fields. This means that the geodesics and the autoparallels in a given
frame depend on the geometric properties of the underlying manifold (fields $g_{ij},$ $%
Q_{.kj}^{i},g_{ij|k})$ and the unholonomicity of the frame (field $%
C_{.jk}^{i})$.

\item The antisymmetric part $\Gamma _{.[jk]}^{i}$ of $\Gamma _{jk}^{i}$
depends only on all fields $Q_{.kj}^{i}$ and $C_{.jk}^{i}.$

\item  The objects of unholonomicity $C_{.jk}^{i}$ behave in the same way as the
components of torsion. This means that even in a Riemannian space where $%
Q_{.kj}^{i}=0,g_{ij|k}=0$ in an unholonomic basis the antisymmetric part $%
\Gamma _{.[jk]}^{i}=-\frac{1}{2}C_{.jk}^{i}\neq 0.$ 
\end{enumerate}

This result has lead to
the misunderstanding that when one works in an unholonomic frame then the torsion is introduced. This statement is clearly not correct. 
This misunderstanding has important consequences because the effects
one will observe in an unholonomic frame will be frame dependent  and
not covariant effects. Therefore all conclusions made in a specific
unholonomic frame must be  restricted to that frame only.

\section{$f(T)$ gravity and cosmology}
With the above considerations in mind, let us consider TEGR and its straightforward extension $f(T)$.
Teleparallelism uses as dynamical objects  the {\it vierbiens}  as unholonomic frames in
spacetime. Following the definitions in the previous section, they  are defined by the requirement $%
g(e_{i},e_{j})=e_{i}.e_{j}=\eta _{ij}$, where $\eta _{ij}={\rm diag}(-1,+1,+1,+1)$
is the Lorentz metric in canonical form. Obviously $g_{\mu \nu }(x)=\eta
_{ij}h_{\mu }^{i}(x)h_{\nu }^{j}(x)$ where $e^{i}(x)=h_{\mu
}^{i}(x)dx^{i}$ is the dual basis. 
Differing from GR,
which uses the torsionless Levi-Civita connection,
Teleparallelism utilizes 
the curvatureless {\it Weitzenb\"{o}ck connection}, whose non-null torsion tensor is defined as
\begin{equation}\label{Wein}
    T_{\mu\nu}^{\beta}
=\hat{\Gamma}_{\nu\mu}^{\beta}
-\hat{\Gamma}_{\mu\nu}^{\beta}
=h_{i}^{\beta}
(\partial_\mu h_{\nu}^{i} - \partial_\nu h^{i}_{\mu}) \;.
\end{equation}
Notice  the 
Ricci rotation coefficients are
$\Omega _{jk}^{i}=T_{jk}^{i}$ and encompass all the 
information concerning the
gravitational field. 
The TEGR Lagrangian for the gravitational field
equations (Einstein equations) is assumed to be:%
\begin{equation}
T={S_{\beta}}^{\mu \nu } {T^{\beta}}_{\mu \nu },  \label{lagrangian}
\end{equation}%
where
\begin{equation}
{S_{\beta}}^{\mu \nu }=\frac{1}{2}({K^{\mu \nu }}_{\beta}+\delta^{\mu} _{\beta
}{T^{\theta \nu }}_{\theta}-\delta^{\nu}_{\beta}{T^{\theta \mu }}_{\theta})  
\label{s}
\end{equation}%
and ${K^{\mu \nu }}_{\beta}$ is the  {\it contorsion}  tensor
\begin{equation}
{K^{\mu \nu }}_{\beta}=-\frac{1}{2}({T^{\mu \nu }}_{\beta}-{T^{\nu \mu} }_\beta
-{T_{\beta}}^{\mu \nu }),  \label{contorsion}
\end{equation}%
which equals the difference of the Levi Civita connection in the holonomic
and the unholonomic frame (see Sec. II for details).

Here,  the gravitational field will be driven by a Lagrangian density
which is a function of the trace $T$. 
Therefore, the corresponding action of $f(T)$ gravity
reads as
\begin{equation}
\label{action}
\mathcal{A}_T=\frac{1}{16\pi G}\int {d^{4}xef(T)}  
\end{equation}%
where $e=det(e_{\mu }^{i}\cdot e_{\nu }^{i})=\sqrt{-g}$. Obviously,
TEGR and thus GR, are restored for $f(T)=T$.

In order to construct a realistic theory of gravity, we have to incorporate 
the matter and radiation fields too. Therefore, the total
action is written as
\begin{eqnarray}
\label{action11}
 A_{\rm tot}= {\cal A}_T+\frac{1}{16\pi G }\int d^4x e
\left(L_m+L_r\right),
\end{eqnarray}
where the matter and radiation Lagrangians  are assumed to correspond
to perfect fluids
with energy densities $\rho_m$, $\rho_r$ and pressures $p_m$, $p_r$
respectively.
If matter couples to the metric in the standard form
then the variation of the action with respect to the vierbein leads to the
equations \cite{Ferraro}
\begin{eqnarray}
&&e^{-1}\partial _{\mu }(e{S}_{i}^{\mu \nu })f^{\prime }(T)-h_{i}^{\lambda }
T_{\mu \lambda }^{\beta}S_{\beta}^{\nu \mu }f^{\prime }(T)  \notag \\
&&+S_{i}^{\mu \nu }\partial _{\mu }(T)f^{\prime \prime }(T)+\frac{1}{4}%
h_{i}^{\nu }f(T)=4\pi Gh_{i}^{\beta}{T^{(m)}}_{\beta}^{\nu }  \label{equations}
\end{eqnarray}%
where a prime denotes differentiation with respect to $T$, ${S_{i}}^{\mu \nu
}={h_{i}}^{\beta}S_{\beta }^{\mu \nu }$ and $T^{(m)}_{\mu \nu }$ is the matter
energy-momentum tensor. It is easy to show that, for $f(T)=T$, Eqs.(\ref{equations}) reduce to the standard Einstein equations \cite{miao}.

In order to consider the related $f(T)$ cosmology,  let us assume a spatially flat 
FLRW metric 
which,
in the holonomic (comoving) frame $\{\partial t,\partial x,\partial
y,\partial z\}$,  assumes  the form
\begin{equation*}
ds^{2}=-dt^{2}+a^{2}(t)(dx^{2}+dy^{2}+dz^{2})
\end{equation*}%
where $a(t)$ is the cosmological scale factor. In this space we define the
vierbein (unholonomic frame) $\{e_{i}\}$ which becomes:
\begin{equation}
h_{\mu }^{i}(t)={\rm diag}(1,a(t),a(t),a(t)),  \label{metric}
\end{equation}
In order to derive the cosmological equations 
in a FLRW metric, we need to deduce a point-like Lagrangian 
from the action (\ref{action}). As a consequence, the infinite degrees of freedom of the original field theory will
be reduced to a finite number as in mechanical systems. This fact allows to deal with minisuperspaces of finite dimensions (see \cite{hamilton} for details). 

In this framework, considering $\{a,T\}$ as the canonical 
variables of the configuration space 
the $f(T)$ action becomes formally:
\begin{equation*}
\mathcal{A}_T=\int \mathcal{L}(a,{\dot{a}},T,\dot{T})dt \;.
\end{equation*}%
Due to the fact that $T$, in GR, reduces to 
\begin{equation}
\label{H.50}
T=-6\left( \frac{\dot{a}}{a}\right)^2 =-6H^{2}  
\end{equation}%
where $H$ is the Hubble parameter \cite{Myrzakulov:2010vz},
 one can rewrite the $f(T)$ action using a Lagrange
multiplier $\lambda_{\cal L}$ as follows:

\begin{equation}
\mathcal{A}_T=2\pi ^{2}\int dt\left\{ f(T)a^{3}-\lambda_{\cal L} \left[ T+6\left( \frac{%
\dot{a}^{2}}{a^{2}}\right) \right] \right\} \,.  \label{H.51}
\end{equation}
In order to determine $\lambda_{\cal L}$, we need to vary the $f(T)$  
action with respect to $T$, that is

\begin{equation*}
a^{3}\frac{df(T)}{dT}\delta T-\lambda_{\cal L} \delta T=0\,
\end{equation*}%
from which follows
\begin{equation*}
\lambda_{\cal L} =a^{3}f^{\prime }(T)\,.
\end{equation*}%
Replacing in the Lagrangian we find:%
\begin{equation}
\mathcal{L}=a^{3}\left[ f(T)-Tf^{\prime }(T)\right] -6\dot{a}^{2}af^{\prime
}(T)\,, 
\label{H.52}
\end{equation}
which is canonical in the variables $\{a,T\}$.

Also, the substitution of the 
vierbein (\ref{metric}) in Eq.(\ref{equations}) for $i=\nu=0 $ (as well as 
the energy condition) yields
\begin{equation}
\label{friedmann}
12H^{2}f^{\prime }(T)+f(T)=16\pi G\rho \;.  
\end{equation}%
Besides, for $i=\nu=1$ Eq.(\ref{equations}) gives
\begin{equation}
48H^{2}\dot{H}f^{\prime \prime }(T)-4(\dot{H}+3H^2)f^{\prime}(T)-f(T)=16\pi Gp
\label{acceleration}
\end{equation}%
where 
$\rho=\rho_{m}+\rho_{r}$ and $p=p_{m}+p_{r}$ are the total
energy density and pressure respectively 
which they have been measured in the
unholonomic frame. 
It is important to stress that Eqs.(\ref{friedmann}), (\ref{acceleration}) can be derived by the Euler-Lagrange equations
\begin{equation}
E_{\mathcal{L}}=\frac{\partial \mathcal{L}}{\partial \dot{a}}\dot{a}+\frac{\partial \mathcal{L}}{\partial \dot{T}}\dot{T}-\mathcal{L}\,,
\end{equation}
and
\begin{equation}
\frac{d}{dt}\frac{\partial \mathcal{L}}{\partial \dot{a}}=\frac{\partial \mathcal{L}}{\partial a}\,,
\end{equation}
respectively. The Euler-Lagrange equation
\begin{equation}
\frac{d}{dt}\frac{\partial \mathcal{L}}{\partial \dot{T}}=\frac{\partial \mathcal{L}}{\partial T}\,,
\end{equation}
gives the constraint (\ref{H.50}). In this sense, the point-like Lagragian  (\ref{H.52}) completely defines the 
related dynamical system in the minisuperspace $\{a,T \}$.

It is interesting to mention that 
using the conservation equation 
$\dot{\rho}+3H(\rho +p)=0$
one can rewrite Eqs. (\ref{friedmann}) and (\ref{acceleration}) in the Friedmann-Einstein 
form 
\begin{equation}
H^{2}=\frac{8\pi G}{3}(\rho +\rho _{T}),  \label{modfri}
\end{equation}%
\begin{equation}
2\dot{H}+3H^{2}=-8\pi G(p+p_{T})  \label{modacce}
\end{equation}%
where
\begin{equation}
\rho _{T}=\frac{1}{16\pi G}[2Tf^{\prime }(T)-f(T)-T],  \label{rhoT}
\end{equation}%
\begin{equation}
p_{T}=\frac{1}{16\pi G}\left\{4\dot{H}[2Tf^{\prime\prime}(T)+f^{\prime}(T)-1]\right\}-\rho_T  \label{pT}
\end{equation}%
are the unholonomicity contributions to the energy density and pressure that disappear as son as $f(T)=T$.
Finally,  $f(T)$ gravity can mimic, under specific circumstances,  
 the scalar field  for dark energy \cite{Myrzakulov:2010vz}. In order to address this crucial
question, we need to derive an effective equation-of-state parameter
$w(a)$ for the $f(T)$ cosmology. 
Indeed, utilizing Eqs. (\ref{rhoT}) and (\ref{pT}), we can 
easily obtain the effective
unholonomicity equation of state as
\begin{equation}
\label{omegaeff}
\omega _{T}\equiv \frac{p_{T}}{\rho _{T}}=-1+\frac{4\dot{H}[2Tf^{\prime
\prime }(T)+f^{\prime }(T)-1]}{2Tf^{\prime }(T)-f(T)-T} \;.  
\end{equation}
It is easy to see that possible deviations from $\Lambda$CDM model can be addressed by the second term in such an equation.

\section{Noether symmetries}
Generally, Noether symmetries play an
important role in physics because they can be used to simplify a given system
of differential equations as well as to determine the integrability of the
system. In general, the existence of a Noether symmetry can be related to a conserved quantity bringing a physical meaning. The so called {\it Noether Symmetry Approach} results extremely useful in cosmology in order to find out exact solutions (see \cite{Cap96} for a comprehensive review of the method).
We would like to
remind the reader that a fundamental approach to derive the Noether 
symmetries for a given dynamical problem (in a 
Riemannian space)
has been published recently by Tsamparlis \& Paliathanasis \cite{Tsam10} (a
similar analysis can be found in \cite{Kalotas,Olver,StephaniB,MoyoLeach,Tsamparlis2010,Tsama10}).

Let us consider the Hamiltonian ${\cal H}$ which 
depends on one independent$~$variable~$\left\{ t\right\} $ and $n$
dependent variables~$\left\{ x^{i}(t):i=1...n\right\} $, i.e. 
${\cal H}={\cal H}\left(
t,x^{k},\dot{x}^{k},...,x^{\left[ n\right] k}\right)$ where a
dot over a symbol means differentiation with respect to $t$. 
We perform the one parameter point transformation
\begin{equation}
\label{Ls.01}
\bar{t}=\Xi \left( t,x^{k},\varepsilon \right)\;,~~\bar{x}^{A}=\Phi \left(
t,x^{k},\varepsilon \right) \;.  
\end{equation}%
In that case, the generating vector of the one parameter point
transformation is
\begin{equation}
X=\xi \left( t,x^{k},\varepsilon \right) \partial _{t}+\eta ^{i}\left(
t,x^{k},\varepsilon \right) \partial _{i}  \label{Ls.02}
\end{equation}%
where%
\begin{equation*}
\xi \left( t,x^{k}\right) =\frac{\partial \Xi ^{i}\left( t,x^{k},\varepsilon
\right) }{\partial \varepsilon }|_{\varepsilon \rightarrow 0}~~,~~\eta
^{i}\left( t,x^{k}\right) =\frac{\partial \Phi \left( t,x^{k},\varepsilon
\right) }{\partial \varepsilon }|_{\varepsilon \rightarrow 0}.
\end{equation*}

The extension of the generator vector in the jet space $B_{M}=\left\{
t,x^{k},\dot{x}^{k},\ddot{x}^{k}...,x^{\left[ n\right] k}\right\} $ is \cite%
{StephaniB}
\begin{equation*}
X^{\left[ n\right] }=X+\eta _{i}^{A}\partial _{u_{i}}+...+\eta
_{ij..i_{n}}^{A}\partial _{u_{ij..i_{n}}}
\end{equation*}%
where%
\begin{equation}
\eta ^{\left[ 1\right] i}=\frac{d}{dt}\eta ^{i}-\dot{x}^{i}\frac{d}{dt}\xi
\end{equation}%
\begin{equation}
\eta ^{\left[ n\right] i}=\frac{d}{dt}\eta ^{\left[ n-1\right] i}-x^{\left[ n%
\right] ^{i}}\frac{d}{dt}\xi
\end{equation}%
$X^{\left[ n\right] }$ is called the nth prolongation of the generator 
(\ref{Ls.02}).

We say that the 
function ${\cal H}\left( t,x^{k},\dot{x}^{k},\ddot{x}^{k}...,x^{\left[
n\right] k}\right) =0$ is invariant under the transformation of Eq.(\ref{Ls.01})
if and only if there is a function $\lambda_{\cal L}$ such as the following
condition holds
\begin{equation}
X^{\left[ n\right] }\left( {\cal H}\right) =\lambda_{\cal L} {\cal H}\;,\;\;mod{\cal H}=0  \label{Ls.03}
\end{equation}%
where $\lambda_{\cal L}$ is a function to be determined 
\cite{IbraB}. Moreover, the
generating vector (\ref{Ls.02}) is a Lie symmetry of 
the function ${\cal H}\left(
t,x^{k},\dot{x}^{k},\ddot{x}^{k}...,x^{\left[ n\right] k}\right)$.
In the following sections we are interested on systems of second order
which implies that the Hamiltonian becomes
${\cal H}={\cal H}\left( t,x^{k},\dot{x}^{k},\ddot{x}^{k}\right)$.

\subsection{Noether Theorems}
Let $\mathcal{L}\left( t,x^{k},\dot{x}^{k}\right)$ be a function which describes
the dynamics of a system. The equations of motion of the dynamical system
follow from the action of the Euler Lagrange vector $E_{i}$ on the function 
$\mathcal{L}$, i.e.%
\begin{equation}
E_{i}\left( L\right) =0.  \label{Ls.04}
\end{equation}%
where the Euler Lagrange vector is
\begin{equation}
E_{i}=\frac{d}{dt}\frac{\partial }{\partial \dot{x}^{i}}-\frac{\partial }{%
\partial x^{i}}
\end{equation}
If the Lagrangian is invariant under the action of the transformation (\ref%
{Ls.01}), namely $X^{\left[ 1\right] }\mathcal{L}=0$ then, it is easy to see
that the Euler Lagrange equations (\ref{Ls.04}) are also invariant under the
transformation (\ref{Ls.01}). In general we have the following theorem.
\cite{StephaniB}

\textbf{\emph{Theorem 1:}}
\textit{Let 
\begin{equation} 
X=\xi \left( t,x^{k}\right) \partial _{t}+\eta ^{i}\left( t,x^{k}\right)
\partial _{i}  \label{Ls.05}
\end{equation}%
be the infinitesimal generator of the transformation (\ref{Ls.01}) and
\begin{equation}
\mathcal{L}=\mathcal{L}\left( t,x^{k},\dot{x}^{k}\right)   \label{Ls.06}
\end{equation}%
be a Lagrangian describing the dynamical system (\ref{Ls.04}). The action of
the transformation (\ref{Ls.01}) on (\ref{Ls.06})  leaves the Euler Lagrange equations (\ref{Ls.04}) invariant, if
and only if there exist a function $g=g\left( t,x^{k}\right) $ such that the
following condition holds
\begin{equation}
X^{\left[ 1\right] }L+L\frac{d\xi }{dt}=\frac{dg}{dt}  \label{Ns.03}
\end{equation}%
where $X^{\left[ 1\right] }$ is the first prolongation of (\ref{Ls.05}).
}

If the generator of Eq.(\ref{Ls.05}) satisfies Eq.(\ref{Ns.03}) then 
the generator (\ref{Ls.05}) is a Noether symmetry of the dynamical system described by the Lagrangian (\ref{Ls.06}). Noether symmetries 
form a Lie algebra called the
Noether algebra. We also have the result

\textbf{\emph{Theorem 2:}}
\textit{For any Noether symmetry (\ref{Ls.05}) of the Lagrangian (%
\ref{Ls.06}) there corresponds a function $I\left( t,x^{k},\dot{x}%
^{k}\right) $
\begin{equation}
I=\xi \left( \dot{x}^{i}\frac{\partial L}{\partial \dot{x}^{i}}-L\right)
-\eta ^{i}\frac{\partial L}{\partial x^{i}}+g  \label{Ls.08}
\end{equation}%
which is a first integral i.e. $\frac{dI}{dt}=0$. The function (\ref{Ls.08})
is called a Noether integral (first integral) of the dynamical system (\ref{Ls.04}).}

\section{Noether Symmetries for  $f\left( T\right) $ cosmology}
In this section we apply the Noether symmetries approach to $f(T)$ 
cosmology in which the corresponding 
Lagrangian of the field equations is given by Eq.(\ref{H.52}).
Here we consider a one parameter point transformation in the space $\left\{
t,a,T\right\}$ and the generator is written as
\begin{equation*}
X=\xi \left( t,a,T\right) \partial _{t}+\eta _{1}\left( t,a,T\right)
\partial _{a}+\eta _{2}\left( t,a,T\right) \partial _{T} \;.
\end{equation*}%
Notice that the Lagrangian (\ref{H.52}) is a singular 
Lagrangian (the Hessian vanishes),
hence the jet space is $\bar{B}_{M}=\left\{ t,a,T,\dot{a}\right\}$
and thus the first prolongation of $X$ in the jet space $\bar{B}_{M}~$is
\begin{equation}
X^{\left[ 1\right] }=\xi \partial _{t}+\eta _{1}\partial _{a}+\eta
_{2}\partial _{T}+\eta _{1}^{\left[ 1\right] }\partial _{\dot{a}}
\end{equation}%
where $\eta _{1}^{\left[ 1\right] }=\dot{\eta}_{1}-\dot{a}\dot{\xi}$
\cite{IbraB,Havelkova, Ziping,Chris2}.
Now we compute each term in the symmetry condition (\ref{Ns.03}).

The term $X^{\left[ 1\right] }L$ gives%
\begin{eqnarray*}
X^{\left[ 1\right] }L &=&\left[ 3a^{2}\eta _{1}\left( f_{T}T-f\right)
+a^{3}f_{TT}T\eta _{2}\right] + \\
&&+\left[ 12f_{T}a\eta _{1,t}\right] \dot{a}+\left[ 12f_{T}a\xi _{,a}\right]
\dot{a}^{3}+ \\
&&+6\left[ f_{T}\eta _{1}+f_{TT}a\eta _{2}+2f_{T}a\eta _{1,a}-2f_{T}a\xi
_{,t}\right] \dot{a}^{2}+ \\
&&+\left[ 12f_{T}a\eta _{1,T}\right] \dot{a}\dot{T}+\left[ 12f_{T}a\xi _{,T}%
\right] \dot{a}^{2}\dot{T}.
\end{eqnarray*}

The second term $L\dot{\xi}$ gives%
\begin{eqnarray*}
L\dot{\xi} &=&\left[ a^{3}\left( f_{T}T-f\right) \xi _{,t}\right] +\left[
a^{3}\left( f_{T}T-f\right) \xi _{,a}\right] \dot{a}+ \\
&&+\left[ a^{3}\left( f_{T}T-f\right) \xi _{,T}\right] \dot{T}+\left[
6f_{T}a\xi _{,t}\right] \dot{a}^{2}+ \\
&&+\left[ 6f_{T}a\xi _{,a}\right] \dot{a}^{3}+\left[ 6f_{T}a\xi _{,T}\right]
\dot{a}^{2}\dot{T}.
\end{eqnarray*}%
Finally the rhs of Eq.(\ref{Ns.03}) is%
\begin{equation*}
\dot{g}=g_{,t}+g_{,a}\dot{a}+g_{,T}\dot{T}.
\end{equation*}

Replacing the results in Eq.(\ref{Ns.03}) and setting the terms with the powers
of $ \dot{a}$   and $\dot{T}$ equal to zero in order to select the Lie vector (see  \cite{Cap96} for details),  
we find the following set of Noether symmetry conditions%
\begin{equation}
\xi _{,a}=0~,~\xi _{,T}=0~,~\eta _{1,T}=0  \label{NC.01}
\end{equation}%
\begin{equation}
a^{3}\left( f_{T}T-f\right) \xi _{,T}=g_{,T}  \label{NC.02}
\end{equation}%
\begin{equation}
3a^{2}\eta _{1}\left( f_{T}T-f\right) +a^{3}f_{TT}T\eta _{2}+a^{3}\left(
f_{T}T-f\right) \xi _{,t}=g_{,t}  \label{NC.03}
\end{equation}%
\begin{equation}
12f_{T}a\eta _{1,t}+a^{3}\left( f_{T}T-f\right) \xi _{,a}=g_{,a}
\label{NC.04}
\end{equation}%
\begin{equation}
f_{T}\eta _{1}+f_{TT}Ta\eta _{2}+2f_{T}a\eta _{1,a}-f_{T}a\xi _{,t}=0
\label{NC.05}
\end{equation}
From equations (\ref{NC.01}), (\ref{NC.02}) it follows%
\begin{equation*}
\xi =\xi \left( t\right) ,~\eta _{1}=\eta _{1}\left( t,a\right) ~,~g=g\left(
t,a\right) .
\end{equation*}%
Then Eq.(\ref{NC.04}) becomes $12f_{T}a\eta _{1,t}=g_{,a}$ Because 
$\eta _{1},g$ are independent of $T$ which follows that
\begin{equation*}
\eta _{1}=\eta _{1}\left( a\right) ~,~g=g\left( t\right) .
\end{equation*}
Dividing Eq.(\ref{NC.05}) with $af_{T}$ we find
\begin{equation}
2\eta _{1,a}+\frac{\eta _{1}}{a}+\frac{f_{TT}}{f_{T}}\eta _{2}-\xi _{,t}=0
\label{NC.05a}
\end{equation}%
from which follows that
\begin{equation*}
\eta _{2}=\frac{f_{T}}{f_{TT}}S\left( a,t\right)
\end{equation*}%
where $S$ is an arbitrary function of its arguments. Taking this result into
consideration the conditions (\ref{NC.03}) and 
(\ref{NC.05a}) become respectively
\begin{equation}
2\eta _{1,a}+\frac{\eta _{1}}{a}+S\left( a,t\right) -\xi _{,t}=0
\label{NC.06}
\end{equation}%
\begin{equation}
3a^{2}\eta _{1}\left( f_{T}T-f\right) +a^{3}f_{TT}T~S+a^{3}\left(
f_{T}T-f\right) \xi _{,t}=g_{,t}.  \label{NC.07}
\end{equation}%
From Eq.(\ref{NC.06}) follows that $S\left( a,t\right) =M\left( a\right)
+N\left( t\right) $ hence we have the final symmetry conditions (where $%
f\neq e^{kT}$ $k=$constant):%
\begin{equation}
2\eta _{1,a}+\frac{\eta _{1}}{a}+M+N-\xi _{,t}=0  \label{NC.06a}
\end{equation}

\begin{equation}
3\frac{\eta _{1}}{a}+\frac{f_{T}T}{f_{T}T-f}~M+\frac{f_{T}T}{f_{T}T-f}%
~~N+\xi _{,t}=\frac{1}{a^{3}\left( f_{T}T-T\right) }g_{,t}.  \label{NC.07a}
\end{equation}
It is obvious that equations (\ref{NC.06}), (\ref{NC.07}) hold for arbitrary 
$f\left( T\right)$ as long as 
$\xi =c_{0}$ and $\eta _{1}=\eta _{2}=0$ (i.e. $S=0$). 
In this case the corresponding Noether integral is the Hamiltonian ${\cal H}$,
implying that the dynamical system is autonomous.
Moreover, the conditions (\ref{NC.06a}),(\ref{NC.07a}) give the following system
of equations%
\begin{equation}
\frac{f_{T}T}{f_{T}T-f}=\frac{n}{n-1}  \label{NC.08}
\end{equation}%
and%
\begin{equation*}
g_{,t}=0~~,~~N=c+\xi _{,t}
\end{equation*}%
\begin{equation*}
2\eta _{1,a}+\frac{\eta _{1}}{a}+M=c
\end{equation*}%
\begin{equation*}
3\frac{\eta _{1}}{a}+\frac{n}{n-1}M=m
\end{equation*}%
\begin{equation*}
\frac{n}{1-n}N-\xi _{,t}=m
\end{equation*}%

Solving the first equation of the system (\ref{NC.08}) we find that
\begin{equation}
f\left( T\right)=f_{0}T^{n}
\end{equation}
where $f_{0}$ is the integration constant. 
In this context we can obtain the Nother symmetries. Specifically, 
in the case of $n\neq \frac{1}{2},\frac{3}{2}$, the Noether symmetry vector is
\begin{eqnarray*}
X_{1} &=&\left( \frac{3C}{2n-1}t\right) \partial _{t}+\left( Ca+c_{3}a^{1-%
\frac{3}{2n}}\right) \partial _{a}+ \\
&&+\left[ \frac{1}{n}\left( \left( c-m\right) n+3c_{3}a^{-\frac{3}{2n}%
}\right) +\frac{3C}{2n-1}+c\right] T\partial _{T}
\end{eqnarray*}%
as well as the corresponding Noether integral is
\begin{equation*}
I_{1}=\left( \frac{3C}{2n-1}t\right) {\cal H}-12f_{0}n\left( Ca^{2}+c_{3}a^{2-\frac{%
3}{2n}}\right) T^{n-1}\dot{a}
\end{equation*}%
where ${\displaystyle C=\frac{m\left( 1-n\right) +nc}{3}}$.

For ${\displaystyle n=\frac{3}{2},}$ the Noether symmetry is given by
\begin{eqnarray}
X_{2} &=&\frac{1}{5}\left( 3c-2m\right) t\partial _{t}+\left[ \left( \frac{c%
}{2}-\frac{m}{6}\right) a+c_{4}\right] \partial _{a}+  \notag \\
&&+\left[ \left( m+11c\right) -\frac{c_{4}}{a}+\frac{2}{5}\left(
8c-2m\right) \right] T\partial _{T}
\end{eqnarray}%
with corresponding Noether integral
\begin{equation*}
I_{2}=\frac{1}{5}\left( 3c-2m\right) t{\cal H}-18f_{0}\left[ \left( \frac{c}{2}-%
\frac{m}{6}\right) a^{2}+c_{4}a\right] T^{\frac{1}{2}}\dot{a} \;.
\end{equation*}%
Finally for ${\displaystyle n=\frac{1}{2},}$ the Noether symmetry becomes
\begin{equation*}
X_{3}=c_{1}t\partial _{t}+\left( -2c_{1}+c_{3}a^{\frac{1}{4}}\right)
\partial _{a}+\left( 4c_{1}+c_{2}+\frac{3c_{3}}{2}a^{-\frac{3}{4}}\right)
T\partial _{T}
\end{equation*}%
and the Noether integral is
\begin{equation*}
I_{3}=c_{1}t{\cal H}-6f_{0}\left( -2c_{1}a+c_{3}a^{\frac{3}{4}}\right) T^{-\frac{1}{%
2}}\dot{a}.
\end{equation*}
We would like to stress that our results are in agreement 
with those of \cite{WeiHao} but they are richer because
we have considered the term $\xi \partial _{t}$ in the generator which is
not done in \cite{WeiHao}.
To this end it becomes evident 
that $f\left( T\right)=f_{0}T^{n}$ is the only form that 
admits extra Noether symmetries 
implying the existence of exact
analytical solutions (see next section).

\section{Exact cosmological solutions}
In this section we proceed in an attempt to analytically solve 
the basic cosmological equations of the 
$f\left( T\right)=f_{0}T^{n}$ gravity model.
In particular from the Lagrangian (\ref{H.52}), we  obtain the main 
field equation 
\begin{equation}
\label{FF}
\ddot{a}+\frac{1}{2a}\dot{a}^{2}+\frac{f^{\prime \prime }}{f^{\prime }}\dot{a%
}\dot{T}-\frac{1}{4}a\frac{f^{\prime }T-f}{f^{\prime }}=0 \;.
\end{equation}
Also differentiating Eq.(\ref{H.50}) we find 
\begin{equation}
\label{FFa}
\dot{T}=12\left[\left( \frac{\dot{a}}{a}\right) ^{3}-\frac{\dot{a}\ddot{a}}{a^{2}} \right] \;.
\end{equation}
Finally, inserting $f\left( T\right)=f_{0}T^{n}$, 
Eq.(\ref{H.50}) and Eq.(\ref{FFa}) into Eq.(\ref{FF}) 
we derive, after some algebra, that
\begin{equation}
\left( 2n-1\right) \left[ \ddot{a}-\frac{\dot{a}^{2}}{2a}\frac{\left(
2n-3\right) }{n}\right] =0
\end{equation}
a solution of which is 
\begin{equation}
\label{aHE}
a(t)=a_{0}t^{2n/3} \;\;\;\;H(t)=\frac{\dot{a}}{a}=\frac{2n}{3t} 
\end{equation}
or
\begin{equation}
\label{HE}
H=H_{0}a^{-3/2n}=H_{0}(1+z)^{3/2n}
\end{equation}
where $n\in {\cal R}^{\star}_{+}-\{\frac{1}{2}\}$, $a(z)=(1+z)^{-1}$ and 
$H_{0}$ is the Hubble constant in agreement with  \cite{WeiHao}.
Also using Eq.\eqref{HE} the
deceleration parameter is given by
\begin{equation}\label{eq:qnu}
q=-1-\frac{{\rm dln}H}{{\rm dln}a}=-1+\frac{3}{2n} \;.
\end{equation}
From Eq.(\ref{aHE}) it is evident that this cosmological model has no 
inflection point. Therefore,
the main drawback of the $f(T)=f_{0}T^{n}$ gravity model
is that the deceleration parameter preserves sign, and therefore the
universe always accelerates or always decelerates
depending on the value of $n$. Indeed, if we consider $n=1$ (TEGR) then
the above solution boils down to 
the Einstein de Sitter model as it should. On the other hand, the 
accelerated expansion of the universe ($q<0$) is recovered for $n>\frac{3}{2}$.
The latter points that 
even if we would admit $n>\frac{3}{2}$ as a mere phenomenological
possibility, we would be also admitting that the universe has been
accelerating forever, which is of course difficult to accept.

Now, we proceed to provide  
the growth factor of the $f(T)=f_{0}T^{n}$.  
In general, the basic equation which governs the evolution
of the matter fluctuations in the linear regime is given by
\begin{equation}
\label{odedelta}
\ddot{\delta}_{m}+ 2H\dot{\delta}_{m}-4 \pi G_{\rm eff} \rho_{m} \delta_{m}=0
\end{equation}
where $\rho_{m}$ is the matter density and
$G_{\rm eff}$ is the effective Newton's parameter which is written as 
\cite{Zheng:2010am}
\begin{eqnarray}
\label{Geff}
G_{\rm eff}=\frac{G}{f^{\prime}(T)} \;.
\end{eqnarray}
Note, that $G$ denotes Newton's gravitational constant. 
On the other hand, using Eqs. (\ref{modfri}) and (\ref{rhoT})
one can easily write 
\begin{eqnarray}
\label{reff}
4\pi G \rho_{m}=\frac{3H^{2}}{2}-4\pi G \rho_{T}=
\frac{3H^{2}}{2}-\frac{2Tf^{\prime}(T)-f(T)-T}{4} \;.
\end{eqnarray}
Therefore, inserting Eqs.(\ref{H.50}), (\ref{Geff}) and (\ref{reff})  
into Eq.(\ref{odedelta}) we have the following general equation 
\begin{equation}
\label{odedelta1}
\ddot{\delta}_{m}+ 2H\dot{\delta}_{m}+\frac{2Tf^{\prime}(T)-f(T)}{4f^{\prime}(T)}\delta_{m}=0 \;.
\end{equation}
We focus now on the $f(T)=f_{0}T^{n}$ gravity model. First of all 
for GR ($n=1$) we have $G_{\rm eff}=G$ and thus, without losing the 
generality,  we can set $f_{0}=1$\footnote{If $f(T)=f_{0}T$ then the 
Newton's constant is just rescaled to be $G_{\rm eff}=G/f_{0}$
which is also constant in time. This result comes directly from the action 
(\ref{action}) (see also \cite{Zheng:2010am}).}. 
Therefore, Eq.(\ref{odedelta1}) becomes
\begin{equation}
\label{odedelta2}
\ddot{\delta}_{m}+ \frac{4n}{3t}\dot{\delta}_{m}-
\frac{2n(2n-1)}{3t^{2}}
\delta_{m}=0 \;.
\end{equation}
Notice, that in order to derive Eq.(\ref{odedelta2}) we have utilized 
Eqs.(\ref{H.50}) and (\ref{aHE}).
Interestingly, the above differential 
equation modifies that of the Einstein de-Sitter model 
in which $n=1$ (GR).
From the mathematical point of view, Eq.(\ref{odedelta2}) is 
of Euler type whose general solution is
\begin{equation}
\delta_{m}(t)=C_{1}t^{2n/3}+C_{2}t^{1-2n}
\end{equation} 
or 
\begin{equation}
\delta_{m}(a)={\tilde C}_{1}a+{\tilde C}_{2}a^{3(1-2n)/2n}
\end{equation} 
where ${\tilde C}_{1}=C_{1}/a_{0}^{3/2n}$ and 
${\tilde C}_{2}=C_{2}/a_{0}^{3(1-2n)/2n}$. In the case of 
$0<n <\frac{1}{2}$ we have 
two growth factors while for $n>\frac{1}{2}$ the only growth factor
is $D_{+}=a \propto t^{2n/3}$. It is interesting to mention 
that if we write the growth factor as a function of the scale factor 
then mathematically it coincides with that of the 
Einstein de-Sitter model \cite{Peeb93}. This result means that the growth 
rate of clustering $f_{+}(a)=d{\rm ln}D_{+}/d{\rm ln}a$ remains constant 
and equal to unity for every
scale factor, implying that the present growth data 
disfavor the $f(T)=f_{0}T^{n}$ gravity. Indeed, in Fig.1 we plot the 
growth data as collected by Basilakos et al. (see \cite{BasNes13} and 
references therein) with the estimated growth rate
function, $f_{+}(z)\sigma_{8}(z)$
[see $f(T)$ - solid line and $\Lambda$CDM - dashed line]. 
Notice, that the theoretical $\sigma_{8}(z)$ is given by 
$\sigma_{8}(z)=\sigma_{8}D_{+}(z)$, where $\sigma_{8}$
is the rms mass fluctuation on
$R_{8}=8 h^{-1}$ Mpc scales at redshift $z=0$.

\begin{figure}
\mbox{\epsfxsize=14.2cm \epsffile{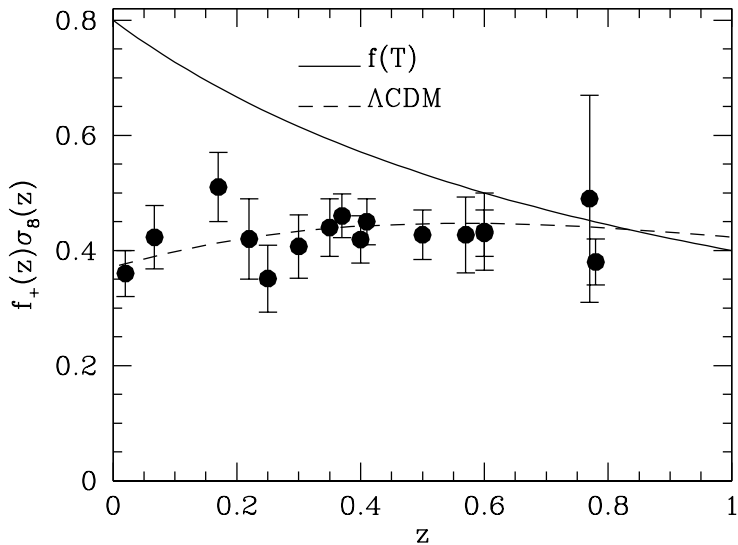}}
\caption{Comparison of the observed (solid points) and
theoretical evolution of the growth
rate $f_{+}(z)\sigma_{8}(z)$. The solid and dashed lines
correspond to $f(T)=f_{0}T^{n}$ and $\Lambda$CDM. 
As in Basilakos et al. \cite{BasNes13} we use 
$\sigma_{8}=0.8$ while for the $\Lambda$CDM case we set $\Omega_{m0}=0.272$.}
\end{figure}

\subsection{Cosmological analogue to other models}
In this section (assuming flatness) we present the cosmological equivalence
at the background level between the
current $f(T)$ gravity with $f(R)$ modified gravity and 
dark energy, through a
specific reconstruction of the $f(R)$ and vacuum energy density namely, 
$f(R)=R^{n}$ and $\Lambda(H)=3\gamma H^{2}$. 
In the case of 
$f(R)=R^{n}$ it has been found by Paliathanasis 
(see Appendix in \cite{Palia12}) that the corresponding 
scale factor obeys Eq.(\ref{aHE}), where 
$n\in {\cal R}^{\star}_{+}-\{2,\frac{3}{2},\frac{7}{8}\}$
\footnote{The Lagrangian here is ${\cal L}_{R}=6naR^{n-1}\dot{a}^{2}+6n(n-1)a^{2}R^{n-2}\dot{a}\dot{R}+(n-1)a^{3}R^{n}$, where $R$ is the Ricci scalar. 
For $n=1$ the solution of the Euler-Lagrange equations is 
the Einstein de-Sitter model [$a(t)\propto t^{2/3}$] as 
it has to be. Note, that 
for $n=2$ one can find a de-Sitter solution 
($a(t) \propto e^{H_{0}t}$, see \cite{Palia12}).}. 
In \cite{prado1,prado2}, it has been shown 
that the particular model $f(R)\propto R^{3/2}$ has the 
cosmological solution ${\displaystyle a(t)=\sqrt{a_4 t^4 + a_3 t^3 + a_2 t^2 + a_1 t}}$ capable of addressing both dark-energy and dark-matter 
dominated phases. However, despite of the analogies, we have 
to point out that $f(R)$ gravity is a fourth-order theory
while $f(T)$ gravity remains of second order.

On the other hand, considering a spatially flat FLRW metric
in the context of GR, the 
combination of the Friedmann equations
with the total (matter+vacuum) energy 
conservation in the matter dominated era provides
(for more details see \cite{BPS09}) 
\begin{equation}\label{LLAA}
{\dot H}+\frac{3}{2}H^{2}=\frac{\Lambda}{2} \;.
\end{equation}
Solving Eq.(\ref{LLAA}) for $\Lambda(H)=3\gamma H^{2}$ 
(see Refs. \cite{FreeseET87,CarvalhoET92,ArcuriWaga94}) we end up with 
\begin{equation}
\label{HE11}
H=H_{0}a^{-3(1-\gamma)/2}=H_{0}(1+z)^{3(1-\gamma)/2} \;.
\end{equation}
Now, comparing Eqs.(\ref{HE}) and (\ref{HE11}) 
and connecting the above coefficients such as $n^{-1}=1-\gamma$,
we find that 
the $f(T)=f_{0}T^{n}$ and the flat $\Lambda(H)=3\gamma H^{2}$ models 
can be viewed as equivalent cosmologies 
as far as the Hubble expansion is concerned, despite the 
fact that the current time varying vacuum model adheres 
to GR.
However, if the $\Lambda(H)=3\gamma H^{2}$ cosmological model
is confronted with the current observations provides a poor fit\,\cite{BPS09}. 
Since the current time varying vacuum model shares 
exactly the same Hubble parameter with the 
$f(T)=f_{0}T^{n}$ gravity model, 
this fact implies that the latter 
is also under observational pressure when we compare against 
the background cosmological data (SnIa, BAOs and CMB data). 
The same observational 
situation holds also for $f(R)=R^{n}$ modified gravity.


\section{Conclusions}
In this paper, we present a general study of Noether symmetries 
for $f(T)$ gravity and discuss the role of torsion and 
unholonomic frames in the context of teleparallel gravity and
its straightforward extension.
In particular, we point out
the misunderstanding that when one works in an unholonomic frame, the 
torsion is introduced showing that this statement is not correct.  
The misunderstanding consists in the fact that the effects
one observes in an unholonomic frame are frame dependent and
not covariant effects. Therefore all conclusions made in a specific
unholonomic frame must be restricted to that frame only. 

Coming to the specific {\it Noether Symmetry Approach}, this article 
extends the works by Basilakos et al. \cite{Basilakos11},
Paliathanasis et al. \cite{Tsafr} and
Wei et al. \cite{WeiHao}. We confirm the result of \cite{WeiHao} that
amongst the variety of $f(T)$ modified
gravity theories, $f(T)=f_{0}T^{n}$ gravity 
admits Noether symmetries (integrals of motion). 
However, we provide here a more
general family of Noether integrals with respect to that of 
\cite{WeiHao}. 
From the mathematical viewpoint the existence of extra integrals of motion
points out the existence of  further analytical solutions.

Based on the $f(T)=f_{0}T^{n}$ models, we derive  
analytical solutions and thus we find the evolution of the main
cosmological functions, namely the scale factor of the universe, the
Hubble parameter, the deceleration parameter and for the first time 
to our knowledge the growth of matter fluctuations in the linear regime. 
Furthermore, we discuss
the linear matter fluctuations from these background solutions.
The analysis of the deceleration parameter points out that the 
$f(T)=f_{0}T^{n}$ gravity models include an intrinsic 
problem namely, the fact that the
expansion of the universe always accelerates or always 
decelerates without spanning the different trends of cosmic evolution.
Another basic problem is related to the fact that the 
growth rate of clustering is constant and always equal to unity 
which means that the present growth data
cannot accommodate the $f(T)=f_{0}T^{n}$ gravity. As shown in \cite{aviles}, a 
robust cosmographic reconstruction of $f(T)$ cosmology needs 
more complicated models to address data. 

Finally, we find 
that flat $f(T)=f_{0}T^{n}$ cosmologically models
are perfectly equivalent to the cosmic
expansion history of the flat $f(R)=R^{n}$ modified gravity and 
the flat time varying vacuum model 
$\Lambda(H)=3\gamma H^{2}$ (where $n^{-1}=1-\gamma$), despite the fact that 
the three models live in a completely different geometrical background. 
This fact is a further indication of the high degeneracy problem 
affecting cosmological models capable of addressing the dark energy issue.


\begin{acknowledgments}
SB acknowledges support by the Research Center for
Astronomy of the Academy of Athens
in the context of the program  ``{\it Tracing the Cosmic Acceleration}''.  SC and MDL are supported by INFN (iniziative specifiche NA12 and OG51).
\end{acknowledgments}


\end{document}